# Approximate Minimum Diameter

Mohammad Ghodsi[1,2], Hamid Homapour[3], and Masoud Seddighin[4]

[1] Sharif University of Technology
[2] IPM - Institute for Research in Fundamental Sciences
ghodsi@sharif.edu
[3] Sharif University of Technology
homapour@ce.sharif.edu
[4] Sharif University of Technology
mseddighin@ce.sharif.edu

**Abstract.** We study the minimum diameter problem for a set of inexact points. By inexact, we mean that the precise location of the points is not known. Instead, the location of each point is restricted to a continues region (*Imprecise* model) or a finite set of points (*Indecisive* model). Given a set of inexact points in one of *Imprecise* or *Indecisive* models, we wish to provide a lower-bound on the diameter of the real points.
In the first part of the paper, we focus on *Indecisive* model. We present an $O(2^{\frac{1}{\epsilon^d}} \cdot \epsilon^{-2d} \cdot n^3)$ time approximation algorithm of factor $(1+\epsilon)$ for finding minimum diameter of a set of points in $d$ dimensions. This improves the previously proposed algorithms for this problem substantially.
Next, we consider the problem in *Imprecise* model. In $d$-dimensional space, we propose a polynomial time $\sqrt{d}$-approximation algorithm. In addition, for $d = 2$, we define the notion of $\alpha$-separability and use our algorithm for *Indecisive* model to obtain $(1+\epsilon)$-approximation algorithm for a set of $\alpha$-separable regions in time $O(2^{\frac{1}{\epsilon^2}} \cdot \frac{n^3}{\epsilon^{10} \cdot \sin(\alpha/2)^3})$.

## 1 Introduction

Rapid growth in the computational technologies and the vast deployment of sensing and measurement tools turned Big Data to an interesting and interdisciplinary topic that attracted lots of attentions in the recent years.

One of the critical issues in Big Data is dealing with uncertainty. Computational Geometry (CG) is one of the fields, that is deeply concerned with large scale data and the issue of imprecision. The real data for a CG problem is often gathered by measurement and coordination instrument (GPS, Digital compass, etc) which are inexact. In addition, the movement is an unavoidable source of inaccuracy. Hence, the data collected by these tools are often noisy and suffer from inaccuracies. In spite of this inaccuracy, most of the algorithms are based on the fact the input data are precise. These algorithms become useless in the presence of inaccurate data. Thus, it is reasonable to design algorithms that take this inaccuracy into account.

Point is the primary object in a geometric problem. Roughly speaking, the input for most of the CG problems is a set of points in different locations of a metric space. The exact location of these points is one of the parameters that might be inaccurate. To capture this inaccuracy in geometric problems, several models have been proposed, e.g. Uncertain, Indecisive, Imprecise and Stochastic models [10]. Suppose that $\mathbb{P} = \{\rho_1, \rho_2, \ldots, \rho_n\}$ is the set of exact points. Due to the measurement inaccuracy, $\mathbb{P}$ is unknown to us. In the following, we give a short description of the way that each model tries to represent $\mathbb{P}$:

- **Uncertain (or locational uncertain) model**: in the *uncertain* model, the location of each point is determined by a probability distribution function. Formally, for every point $\rho_i \in \mathbb{P}$ a probability distribution $\mathcal{D}_i$ is given in input, that represent the probability that $\rho_i$ appears in every location.
- **Indecisive model**: in the *indecisive* model, each input point can take one of $k$ distinct possible locations. In other words, for each real point $\rho_i \in \mathbb{P}$, a set of $k$ points is given, where the location of $\rho_i$ equals to one of these points.
- **Imprecise model**: in the *imprecise* model, possible locations of a point is restricted to a region, i.e., for each point $\rho \in \mathbb{P}$, a finite region $R_i$ is given in input. All we know is $\rho_i \in R_i$.
- **Stochastic model**: every point $p$ in input has a deterministic location, but there is a probability that $p \notin \mathbb{P}$. *Stochstic* model is used for the case that false reading is possible, usually in database scenarios [2, 4].

Let $C$ be an input related to one of these models, that represents the exact point set $\mathbb{P}$. The question is, what information can we ritrieve from $C$? For brevity, suppose that the objective is to compute a statistical value $D(\mathbb{P})$ (e.g., $D(\mathbb{P})$ can be the diameter or width of $\mathbb{P}$). Trivially, finding the exact value of $D(\mathbb{P})$ is not possible, since we don't have access to the real points $\mathbb{P}$. Instead, there may be several approaches, based on the input model. For example, one can calculate the distribution of $D(\mathbb{P})$ (for uncertain model), or can provide upper bound or lower bound on the value of $D(\mathbb{P})$ (for Indecisive and Imprecise models) [10,12].

In this paper, we investigate on *Indecisive* and *Imprecise* models and consider the diameter as the objective function. Furthermore, our approach is to compute bounds on the output. Thus, the general form of the problem we wish to observe is as follows: Given an input $C$ corresponding to an *Indecisive* (*Imprecise*) data model of a set $\mathbb{P}$ of exact data. The goal is to lower-bound the value of the $D(\mathbb{P})$, i.e., finding a lower-bound on the minimum possible value for $D(\mathbb{P})$. In section 1.1 you can find a formal definition of the problem. It is worth mentioning that this problem has many applications in computer networking and databases [7, 8, 15].

### 1.1 Model definition and problem statement

**Indecisive model.** As mentioned, in *Indecisive* model, each actual point $\rho_i \in \mathbb{P}$ is represented via a finite set of points with different locations. For simplicity, suppose that we color the representatives of each point with a unique color. Thus, the input for a problem in *Indecisive* model is a set $\mathcal{P} = \{\mathcal{P}_1, \mathcal{P}_2, \ldots, \mathcal{P}_m\}$, where each $\mathcal{P}_i$ is a set of points representing alternatives of $\rho_i$, i.e., $\rho_i$ is one of the elements in $\mathcal{P}_i$. All the points in $\mathcal{P}_i$ are colored with color $\mathcal{C}_i$. For Indecisive model, we assume that total number of the points is $n$, i.e. $\sum_i |\mathcal{P}_i| = n$.

In addition to uncertainty, this model can represent various situations. For example, consider an instance of a resource allocation problem where we have a set of resources and each resource has a number of copies (or alternatives) and the solutions are constrained to use exactly one of the alternatives. Indecisive model can be used to represent this kind of issues. To do this, each resource can be represented by a point in $d$ dimensional Euclidean space. To represent the type of resources, each point is associated with a color indicating its type.

**Imprecise model.** In *Imprecise* model, the possible locations of a point is restricted by a finite and continuous area. The input for an imprecise problem instance is a set $\mathcal{R} = \{\mathcal{R}_1, \mathcal{R}_2, \ldots, \mathcal{R}_n\}$, where for every region $\mathcal{R}_i$, we know that $\rho_i \in \mathcal{R}_i$. Therefore, each point of $\mathcal{R}_i$ is a possible location for the actual point $\rho_i$.

This model can be applied in many situations. For example, we know that all measurements have a degree of uncertainty regardless of precision and accuracy. To represent an actual point, we can show the amount of inaccuracy with a circle (i.e. as a region) which is centered at the measured point with a radius equals to the size of tolerance interval in the measurement tool, which means that the actual point lies somewhere in this circle.

**Problem statement** For brevitiy, we use $MinDCS$ and $MinDiam$ to refer to the problem in models *Indecisive* and *Imprecise*, respectively. In the following, we formally define $MinDCS$ and $MinDiam$ problems.

**Problem 1** ($MinDCS$) Given a set $\mathcal{P} = \{\mathcal{P}_1, \mathcal{P}_2, \ldots, \mathcal{P}_m\}$, with each $\mathcal{P}_i$ being a finite set of points in $d$-dimensional euclidean space with the same color $\mathcal{C}_i$ and $\sum_i |\mathcal{P}_i| = n$. A *color selection* of $\mathcal{P}$ is a set of $m$ points, one from each $\mathcal{P}_i$. Find a color selection $S$ of $\mathcal{P}$ so that the diameter of these points is the smallest among all options.

Denote by $D_{min}$, the diameter of the desired selection in $MinDCS$. Formally:

$$D_{min} = \min_{S=Sel(\mathcal{P})} \left( \max_{\forall p_i, p_j \in S} ||p_i - p_j|| \right),$$

where $Sel(\mathcal{P})$ is the set of all possible color selections of $\mathcal{P}$. Furthermore, we define $OPT$ as the color selection that the diameter of its points is smallest among all possible color selections:

$$OPT = \arg \min_{S=Sel(\mathcal{P})} \left( \max_{\forall p_i, p_j \in S} ||p_i - p_j|| \right).$$

We also denote by $D(\mathcal{Y})$, the diameter of the point set $\mathcal{Y}$.

**Problem 2** ($MinDiam$) In this problem, a set $\mathcal{R} = \{\mathcal{R}_1, \mathcal{R}_2, \ldots, \mathcal{R}_n\}$ is given, where each $\mathcal{R}_i$ is a bounded region in $d$-dimensional euclidean space. We want to select one point from each region, such that the diameter of the selected points is minimized. Formally, we want to select a set $S = \{p_1, p_2, \ldots, p_n\}$ of points, where $p_i \in \mathcal{R}_i$ such that $D(S)$ is the smallest possible.

### 1.2 Related works

For *Indecisive* model, Fan et al. [11] suggested a randomized algorithm with the time complexity $O(n^{1+\epsilon})$ for $(1+\epsilon)$-approximation of the maximum diameter, where $\epsilon$ could be an arbitrarily small positive constant.

Zhang et al. [15] suggested an $O(n^k)$ time brute force algorithm to find minimum possible diameter in *Indecisive* model. Furthermore, Fleischer and Xu [7,8] showed that this problem can be solved in polynomial time for the $L_1$ and $L_\infty$ metrics, while it is NP-hard for all other $L_p$ metrics, even in two dimensions. They also gave an approximation algorithm with a constant approximation factor. By extending the definition of $\epsilon$-kernels from Agarwal et al. [1] to avatar $\epsilon$-kernels (the notion of avatar is the same as *Indecisive* model) in $d$ dimensional Euclidean space, Consuegra et al. [3] proposed an $(1+\epsilon)$-approximation algorithm with running time $O((n)^{(2d+3)} \cdot \frac{m}{\delta^d} \cdot (2d)^2 \cdot 2^{\frac{1}{\delta^d}}(\frac{2}{\delta^{d-1}})^{\lfloor \frac{d}{2} \rfloor} + (\frac{2}{\delta^{d-1}})^a)$ (where $\delta$ is the side length of the grid cells for constructing core-set, $d$ is the dimension of the points, $k$ is the maximum frequency of the alternatives of a points, and $a$ is a small constant). Furthermore, for *Uncertain* model, Fan et al. [6] proposed an approximation algorithm to compute the expected diameter of a point set. Several other objectives are considered for the *Indecisive* model, i.e., unit covering [5].

In *Imprecise* model, in the Euclidean plane, Löffler and van Kreveld [13] proposed an $O(n \log n)$ time algorithm for the minimum diameter of a set of imprecise points modeled as squares and a $(1+\epsilon)$-approximation algorithm with running time $O(n^{c\epsilon^{-\frac{1}{2}}})$ for the points modeled as discs, where $c = 6.66$.

### 1.3 Our results and techniques

The first part of our work (Section 2) is devoted to the diameter problem in *Indecisive* model. In this case, we present an $O(2^{\frac{1}{\epsilon^d}} \cdot \frac{1}{\epsilon^{2d}} \cdot n^3)$ time approximation algorithm of factor $(1+\epsilon)$. This improves the previous $O((n)^{(2d+3)} \cdot \frac{m}{\delta^d} \cdot (2d)^2 \cdot 2^{\frac{1}{\delta^d}}(\frac{2}{\delta^{d-1}})^{\lfloor \frac{d}{2} \rfloor} + (\frac{2}{\delta^{d-1}})^a)$ time algorithm substantially. The idea is to build several grids with different side-lengths and round the points in $\mathcal{P}$ to the grid points.

In the second part of our work (Section 3) we study the problem in *Imprecise* model. In $d$-dimensional space, we propose a polynomial time $\sqrt{d}$-approximation algorithm. We obtain this result by solving a linear program that formulates a relaxed version of the problem. Next, for $d = 2$, we define the notion of $\alpha$-separability and use a combination of our previouse methods to obtain a polynomial time $(1+\epsilon)$-approximation algorithm for $\alpha$-separable regions. For this purpose, we first determine a lower-bound on the minimum diameter by the $\sqrt{d}$ approximation algoithm, and convert the problem into an *Indecisive* instance. Next, we use the results for *Indecisive* model to obtain a $(1+\epsilon)$-approximation.

## 2 Approximation algorithm for *MinDCS*

We use a powerful technique that is widely used to design approximation algorithms for geometric problems. The technique is based on the following steps [1, 9, 14]:

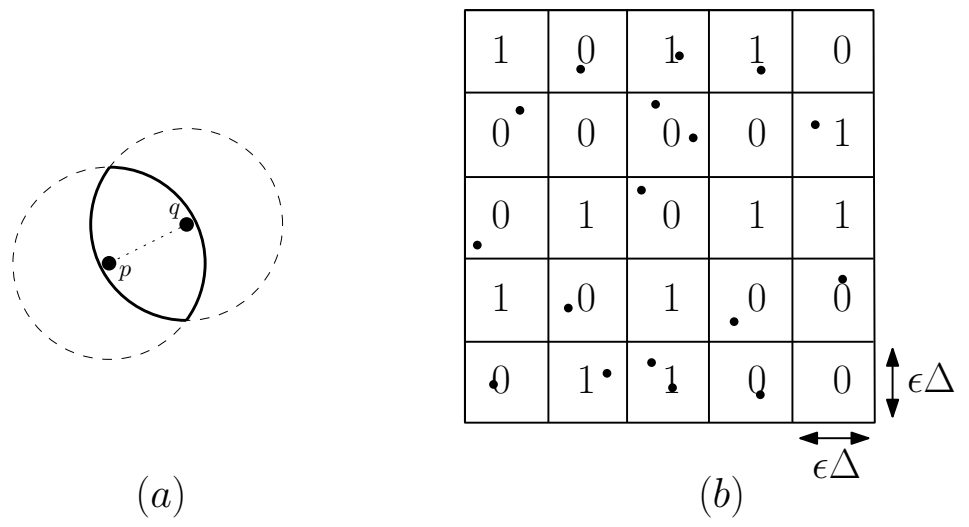

(a) (b)

**Fig. 1.** (a) The possible area of two points $p$ and $q$. (b) A binary assignment for a grid.

1. Extract a small subset of data that inherited the properties of original data (i.e., coreset)
2. Run lazy algorithm on the coreset.

For a set $P$ of points in $R^d$, and an optimization problem $\mathcal{X}$ (e.g., $\mathcal{X}(P)$ would be the diameter or width of $P$), a subset $Q$ of the points of $P$ is an $\epsilon$-coreset, if

$$(1-\epsilon)\mathcal{X}(P) \leq \mathcal{X}(Q) \leq (1+\epsilon)\mathcal{X}(P).$$

We state this fact, by saying that $Q$ is an $\epsilon$-coreset of $P$ for $\mathcal{X}(.)$. Considering the input data in indecisive model, we have the following definition of coreset: given a set $P$ of $n$ points colored with $m$ colors. We say $Q$ is an $\epsilon$-coreset of $P$ for $MinDCS$ iff: $(i)$ $Q$ contains at least one point of each color, $(ii)$ $D(Q) \leq (1+\epsilon)D_{min}$.

### 2.1 Approximate minimum diameter.

**Definition 1.** *Given a set of colored points $P$ and a set $C$ of colors. $P$ is $C$-legal iff for each $c \in C$ there exists a point $p \in P$ with color $c$.*

**Definition 2 (Possible area).** *Consider two points $p$ and $q$. Draw two balls of radius $|pq|$, one of them centered at $p$ and the other centered at $q$. Name the intersection area of these two balls as possible area ($C_{pq}$), see Fig. 1(a).*

**Observation 1** *If $p$ and $q$ be the points that determine the diameter of a point set, all the points in the set must lay in $C_{pq}$.*

Regarding Observation 1, we compute an $\epsilon$-approximation of $MinDCS$ by the process described in Algorithm 1. The algorithm operates as follows: let $\mathcal{S} = \{\mathcal{S}_1, \mathcal{S}_2, \ldots\}$ be the set of all pairs of points in $\mathcal{P}$. For each $\mathcal{S}_i = \{p_i, q_i\}$, let $P_i$ be the set of points in $C_{p_i q_i}$. For each $P_i$ which is $C$-legal, we compute an approximation of minimum diameter of $P_i$ by Algorithm 2, as will be described further. Next, among all computations, we choose the one with the minimum value ($D_{alg}$).

Note that, since we consider all pairs of the points, for some $C$-legal pair $\mathcal{S}_i$ we have $D_{min} = |p_i q_i|$ and hence $D_{alg}$ is an $\epsilon$-approximation of $D_{min}$.

**Algorithm 1** Minimum Diameter Approximation

**function** MinDiameterAPX($\mathcal{P}$)
    $D_{alg} = null$
    **for each** $\mathcal{S}_i \in \mathcal{S}$ **do**
        **if** $P_i$ is $C$-legal **then**
            $D_{apx}(P_i) = DiameterAPX(P_i, \{p_i, q_i\})$
            $D_{alg} = \min\{D_{alg}, D_{apx}(P_i)\}$
        **end if**
    **end for**
    **return** $D_{alg}$
**end function**

In Algorithm 2, you can find the procedure for finding an approximation of minimum diameter for each $P_i$.

The description of Algorithm 2 is as follows: let $\Delta = |p_i q_i|$. First, we compute the smallest axis parallel bounding box of $P_i$ $(B(P_i))$. Next, we split $B(P_i)$ into the cells with side lengths $\epsilon\Delta$ and name the produced uniform grid as $\mathbb{G}$. A *binary assignment* of $\mathbb{G}$ is to assign 0 or 1 to each cell of $\mathbb{G}$, see Fig. 1(b). Consider all binary assignments of $\mathbb{G}$. In the $j$th assignment, let $Q_j$ be the set of the cells with value '1'. We call $Q_j$ *legal*, if the set of points in $Q_j$'s cells is $C$-legal.

Number of the cells in $\mathbb{G}$ is $O(\frac{1}{\epsilon^d})$ and hence, there are at most $O(2^{\frac{1}{\epsilon^d}})$ legal assignments. For each cell of a legal $Q_j$, choose an arbitrary point in that cell as a representative. Next, we compute the diameter of the representatives in time $O((\frac{1}{\epsilon^d})^2)$. Regarding all the computations, we return the minimum of them as an approximation of $D(P_i)$.

Note that if $S_i$ would be the optimum color selection of $P_i$, then obviously there exists an assignment $j$, such that $Q_j$ is legal and includes the cells corresponding to the points in $S_i$.

In order to determine whether or not $Q_j$ is legal, we can check in $O(n)$ time the existence of each color in at least one cell of $Q_j$.

Finally, for all $C$-legal $P_i$, we select the smallest among all approximations of $P_i$ as an approximation of $D_{min}$.

**Theorem 2.** *The MinDCS problem can be approximated in time* $O(2^{\frac{1}{\epsilon^d}}.\epsilon^{-2d} .n^3)$ *of factor* $(1+\epsilon)$ *for fixed dimensions.*

*Proof.* Let $D_{alg}$ be the value returned by our algorithm, and $p, q \in OPT$ be the points with maximum distance in the optimal solution. Obviously, the set of points in $C_{pq}$ is $C$-legal, and all the points in $OPT$ are in $C_{pq}$. Consider the binary assignment correspondinh to $OPT$, i.e., a cell is 1, iff it contains at least one point of $OPT$. Since this assignment is $C$-legal, it would be considered by our algorithm. Thus, $D_{alg} \leq (1+\epsilon)D(OPT)$. On the other hand, the assignment related to $D_{alg}$ is also $C$-legal. Hence, $D_{alg} \geq (1-\epsilon)D(OPT)$.

There are $n^2$ different possible areas. For each of them we have $2^{\frac{1}{\epsilon^d}}$ different assignments. Checking the legality of each assignment takes $O(n)$ time. For a

**Algorithm 2** Approximate $D_{P_i}$ Respect to $p, q \in S_i$

```
function DiameterAPX(P_i, S_i)
    Let S_i = {p_i, q_i}
    Δ = |p_i q_i|
    D(P_i) = null
    Let G be a uniform grid on B(P_i) in d dimensional space with cells of size εΔ
    for each binary assignment of cells of G do
        Let Q_j be the cells that is assigned a value of 1
        if Q_j is legal then
            Let Q'_j be the set of representative points of cells
            D(P_i) = min{D(P_i), D(Q'_j)}
        end if
    end for
    return D(P_i)
end function
```

$C$-legal assignment we can find the minimum diameter in $O(\epsilon^{-2d})$ time since each assignment contains at most $\epsilon^{-d}$ cells. Thus, total running time would be $O(2^{\frac{1}{\epsilon^d}}.\epsilon^{-2d}.n^3)$.

It is worth mentioning that we can improve the running time to $O(2^{\frac{1}{\epsilon^d}} \cdot (\epsilon^{-2d} + n^3))$ by a preprocessing phase that computes the diameter for every $2^{\frac{1}{\epsilon^d}}$ different binary assignments and uses these preprocessed values for every pair of $S$. .tex

## 3 Approximation algorithm for *MinDiam*

In this section, we consider the problem of finding minimum diameter in $Imprecise$ model. As mentioned, in this model an imprecise point is represented by a continues region. Given a set $\mathcal{R} = \{\mathcal{R}_1, \mathcal{R}_2, \ldots, \mathcal{R}_n\}$ where each $\mathcal{R}_i$ is a polygonal region. A **selection** $S = \{s_1, s_2, ..., s_n\}$ of $\mathcal{R}$ is a set of points where $s_i \in \mathcal{R}_i$. In the minimum diameter problem, the goal is to find a selection $S$ such that $D(S)$ is minimized among all options. We consider the problem for the case where every $\mathcal{R}_i$ is a polygonal convex region with constant complexity (i.e., number of edges).

In Section 3.1, we present a polynomial time $\sqrt{d}$-approximation algorithm for $MinDiam$.

### 3.1 $\sqrt{d}$-approximation

Our $\sqrt{d}$-approximation construction is based on $LP$-programming which uses the rectilinear distances ($d_{l_1}$) of the points. For now, suppose that $d$ is constant and consider $LP$ 1. In this $LP$, we want to select the points $s_1, s_2, \ldots, s_n$ such that $s_i \in \mathcal{R}_i$ and the rectilinear diameter of these points is minimized.

$$\begin{array}{lll} \textbf{LP 1:} & & \\ \text{minimize} & \ell & \\ \text{subject to} & d_{l_1}(s_i,\ s_j) \leq \ell & \forall i, j \\ & s_i \in \mathcal{R}_i & \forall i, j \end{array} \tag{1}$$

We show that $d_{l_1}(s_i, s_j)$ can be expressed by $2^d$ linear constraints (note that $d_{l_1}(s_i, s_j) = |s_{i,1} - s_{j,1}| + \ldots + |s_{i,d} - s_{j,d}|$). Based on the relative value of $s_{i,k}, s_{j,k}$, there are $2^d$ different possible expressions for $d_{l_1}(s_i, s_j)$. For example, for $d = 2$, we have:

$$d_{l_1}(s_i,\ s_j) = \begin{cases} s_{i,1} - s_{j,1} + s_{i,2} - s_{j,2} & s_{i,1} \geq s_{j,1} \quad s_{i,2} \geq s_{j,2} \\ s_{j,1} - s_{i,1} + s_{i,2} - s_{j,2} & s_{i,1} < s_{j,1} \quad s_{i,2} \geq s_{j,2} \\ s_{i,1} - s_{j,1} + s_{j,2} - s_{i,2} & s_{i,1} \geq s_{j,1} \quad s_{i,2} < s_{j,2} \\ s_{j,1} - s_{i,1} + s_{j,2} - s_{i,2} & s_{i,1} < s_{j,1} \quad s_{i,2} < s_{j,2} \end{cases}$$

Call these expressions as $e_1, e_2, \ldots, e_{2^d}$. Note that if $d_{l_1}(s_i,\ s_j) \leq \ell$, then for all $1 \leq k \leq 2^n$, $e_k \leq \ell$. With this in mind, we can replace $d_{l_1}(s_i,\ s_j) \leq \ell$ with these $2^d$ linear expressions. It is worth mentioning that we can reduce the number of expressions of this type in *LP* 1 using some additional variables, see Section 3.2 for more details.

Since the regions are convex, the constraint $s_i \in \mathcal{R}_i$ can be described by $|\mathcal{R}_i|$ linear inequalities where $|\mathcal{R}_i|$ is the complexity of region $\mathcal{R}_i$. Thus, the number of the constraints of this type is $|\mathcal{R}| = \sum_i |\mathcal{R}_i|$. Therefore, total number of the constraints would be $2^d \cdot n^2 + |\mathcal{R}|$ which is $\mathsf{poly}(n)$ for constant $d$.

**Theorem 3.** *$D_{min} \leq \ell \leq \sqrt{d}D_{min}$.*

*Proof.* Let $S_{min}$ be the optimal selection of $R$ so that $D_{min} = D(S_{min})$. Obviously, $\ell \geq D_{min}$. On the other hand, $S_{min}$ is a feasible solution for *LP* 1 with rectilinear diameter at most $\sqrt{d}D_{min}$.

Theorem 3 shows that *LP* 1 results in a $\sqrt{d}$ approximation algorithm for *MinDiam*. We use this result to bound the optimal solution in Section 3.3.

### 3.2 Reducing the number of constraints

In this section, we describe how to reduce the number of constraints in *LP* 1. For this, we use Observation 4.

**Observation 4** *$|x|$ is the smallest value $z$ such that $z \geq x$ and $z \geq -x$.*

Thus, for the following *LP*, we have $z = |x|$:

$$\begin{array}{ll} \textbf{LP 2:} & \\ \text{minimize} & z \\ \text{subject to} & x \leq z \\ & -x \leq z \end{array} \tag{2}$$

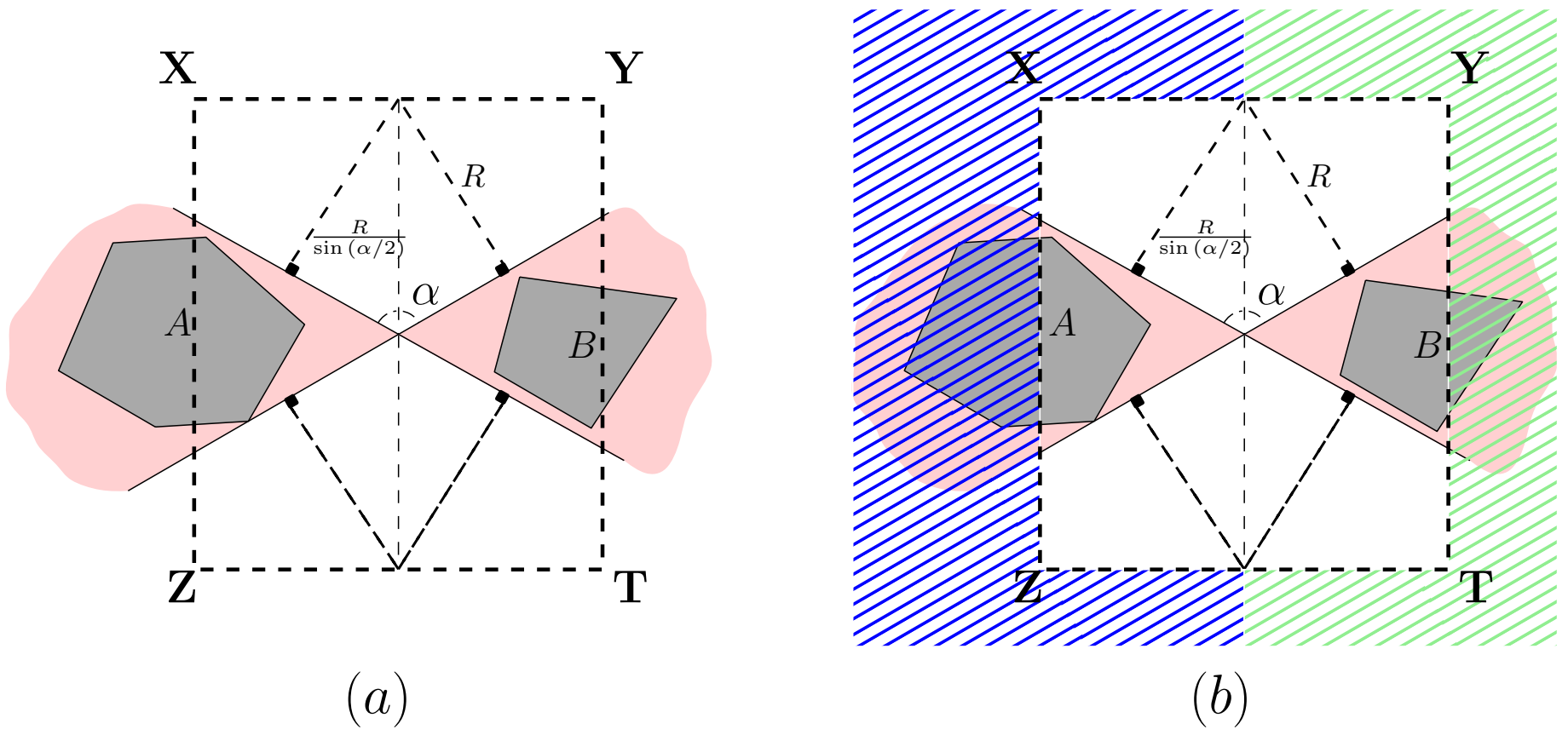

**Fig. 2.** (a) Regions are separated by two intersecting lines. The optimal solution $S_{min}$, must be entirely within the rectangle $XYZT$. (b) The points inside the green area are at least $R$ distance away from the region $A$ (similar fact is intact between the points inside the blue area and the region $B$).

We use Observation 4 to modify $LP$ 1 into $LP$ 3. In $LP$ 3, $Q$ is a large enough integer, ensuring that minimizing $\ell$ is more important than $\sum d_{i,j,k}$.

**LP 3:**

$$\begin{array}{lll} \text{minimize} & Q\ell + \sum d_{i,j,k} & \\ \text{subject to} & \sum_k d_{i,j,k} \leq \ell & \forall i,j \\ & s_i \in \mathcal{R}_i & \forall i,j \\ & d_{i,j,k} \geq s_{i,k} - s_{j,k} & \forall i,j,k \\ & d_{i,j,k} \geq s_{j,k} - s_{i,k} & \forall i,j,k \end{array} \tag{3}$$

Number of the constraints in $LP$ 3 is $O(n^2 d + |\mathcal{R}|)$ which is $\mathsf{poly}(n)$.

### 3.3 $(1+\epsilon)$-approximation for $d = 2$

In this section, we propose an $\epsilon$-approximation algorithm for $MinDiam$ when $d = 2$ and the regions in $\mathcal{R}$ admit a special notion of separability, called $\alpha$-separability.

**$\alpha$-separability**

**Definition 3.** *Two intersecting lines, divide the plane into four areas. Two regions $A$ and $B$ are separated by these lines, if they completely belong to opposite areas, See Fig. 2(a). In Fig. 2(a), we name $\alpha$ as the degree of separation.*

**Definition 4.** *A set $S$ of regions is $\alpha$-separable, if **there exists** two regions $X$ and $Y$ in $S$, which are $\alpha$-separable.*

**Definition 5.** *For two regions $X$ and $Y$, maximum degree of separability is defined as the maximum $\alpha$, such that $X$ and $Y$ are $\alpha$-separable. Subsequently, maximum degree of separability for a set $\mathcal{R}$ of regions, is the maximum possible $\alpha$, such that $\mathcal{R}$ is $\alpha$-separable.*

It's easy to observe that the maximum degree of separability for two regions can be computed in polynomial time by computing the common tangent of the regions. Similarly, maximum degree of separability for a set $\mathcal{R}$ of regions, can be computed in polynomial time.

Note that the maximum degree of separability is not well defined for the case, where every pair of regions have a nonempty intersection. We postpone resolving this issue to Section 3.4. For now, we suppose that $\mathcal{R}$ contains at least two regions that are completely disjoint.

### $(1+\epsilon)$-approximation algorithm

**Theorem 5.** *Let $\mathcal{R}$ be an $\alpha$-separable set of regions. then there exists an $\epsilon$-approximation algorithm for computing MinDiam of $\mathcal{R}$ with running time $O(2^{\frac{1}{\epsilon^2}} \cdot \frac{n^3}{\epsilon^{10}.\sin(\alpha/2)^3})$.*

*Proof.* Assume that $A, B$ are the regions in $\mathcal{R}$, that are $\alpha$-separable. Fig. 2(a) shows the regions and the separating lines. Let $R$ be the value, with the property that $D_{min} \leq R \leq \sqrt{2}D_{min}$. Such $R$ can be computed by the $\sqrt{d}$-approximation algorithm described in section 3.1.

Argue that the points in the optimal solution $S_{min}$ must be entirely within the rectangle $XYZT$ (See Fig. 2(a)), while every point out of the rectangle has distance more than $R$ to at least one of the selected points from regions $A$ or $B$ (see Fig. 2(b)). The area of $XYZT$ is $\frac{4R^2}{\sin(\alpha/2)}$, that is, $O(\frac{R^2}{\sin(\alpha/2)})$.

Color each region $R_i$ with color $c_i$. Now, construct a grid $\mathbb{G}$ on the rectangle $XYZT$ with cells of size $\epsilon R$. For each grid point $p$ and region $R_i$ containing $p$ create a point with the same location as $p$ and color $c_i$, see Fig. 3(a). Total number of the points is $O(\frac{n}{\epsilon^2 \sin(\alpha/2)})$.

Let $P$ be the set of generated points. We give $P$ as an input instance for $MinDCS$ problem. Next, using algorithm 1 we find a color selection $Q$ with $D(Q) \leq (1+\epsilon)D_{min}$ where $D_{min}$ is the diameter of the optimal selection. While the diameter of each cell in $\mathbb{G}$ equals $\sqrt{2}\epsilon = O(\epsilon)$, total error of selection $Q$ for *MinDiam* is $O(\epsilon)$. Considering the time complexity obtained for $MinDCS$, total running time would be $O(2^{\frac{1}{\epsilon^2}} \cdot \frac{n^3}{\epsilon^{10} \cdot \sin(\alpha/2)^3})$.

Note that $1/\sin(\alpha)$ is exponentially descending for $0 \leq \alpha \leq \pi/2$. As an example, for $\alpha > 20°$, $1/\sin(\alpha) < 3$ and for $\alpha > 45°$, $1/\sin(\alpha) < 1.5$. Thus, for large enough $\alpha$ (e.g. $\alpha > 10$), we can consider this value as a constant.

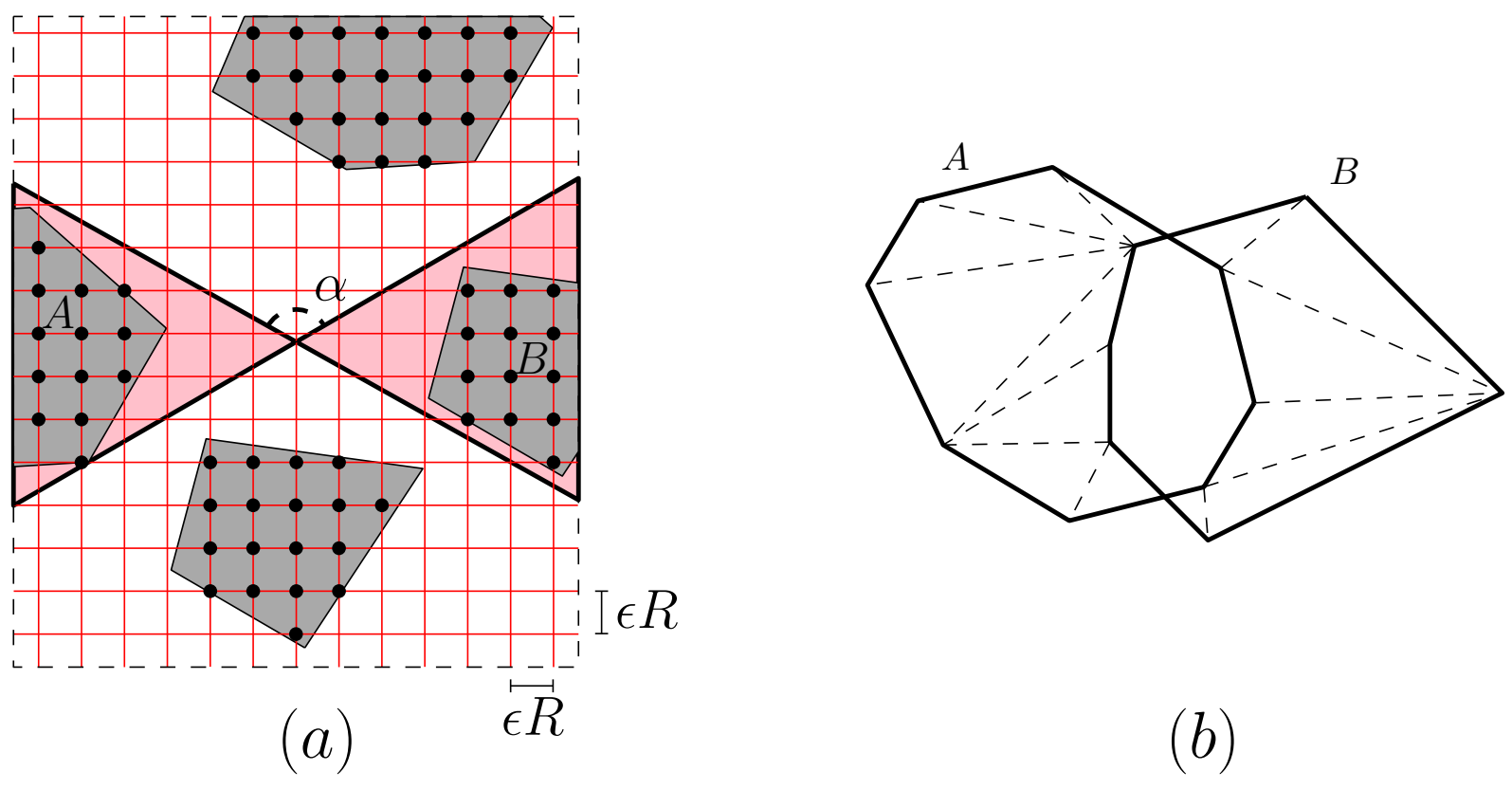

**Fig. 3.** (a) Constructing a grid on $XYZT$. (b) Special cases.

### 3.4 A discussion on $\alpha$-separability

As previously mentioned, $\alpha$-separability is not well defined in the case that all pairs of the regions have a nonempty intersection. In this section, we wish to address this issue.

Note that by **Helly's theorem**, if every triple regions of the set $\mathcal{R}$ have a nonempty intersection, then the intersection of all the regions in $\mathcal{R}$ is nonempty, i.e., $\bigcap_i \mathcal{R}_i \neq \emptyset$. In this case, the optimal selection in $MinDiam$ is trivial: select the same point for all the regions. Thus, we can assume that there exists a set of three regions, namely, $A, B$, and $C$ in $\mathcal{R}$ such that any pair of them has a nonempty intersection, but $A \cap B \cap C$ is empty.

Denote by $p_A, p_B$, and $p_C$ the selected points from the regions $A, B$, and $C$, respectively. To solve the problem of $MinDCS$ on $\mathcal{R}$ we divide the problem into three sub problems: $(i)$ $\{p_A, p_B\} \in A \cap B$, $(ii)$ $p_A \in (A \setminus B)$ and $p_B \in (B \setminus A)$, and $(iii)$ $p_A \in A \cap B$ and $p_B \in (A \setminus B)$.

For the first case, we can assume w.l.o.g that $p_A$ and $p_B$ are in the same location in $A \cap B$. Thus, we can remove $A \Delta B$ and only keep $A \cap B$. Since the intersection of $(A \cap B)$ and $C$ is empty, they are separable.

For the second case, we partition $A \Delta B$ into the constant number of convex sub-regions (for example, by triangulating $A \Delta B$ as shown in Fig. 3(b)). Suppose that $A \setminus B$ is divided into $r$ sub-regions and $B \setminus A$ is divided into $s$ sub-regions. Then we solve $r \times s$ sub-problems, according to the sub-regions $p_A$ and $p_B$ belong. In all of these sub-problems, $\alpha$-separability is well defined.

Finally, note that considering the third case is not necessary, since if $p_A \in A \cap B$ and $p_B \notin A \cap B$ then we can move $p_B$ to the same location as $p_A$ without any decrease in the diameter of the selected points.

.tex

## 4 Conclusions and Future Works

In this paper, we tried to address the diameter problem in two models of uncertainty. In section 2, we investigate on the problem of minimum diameter in *Indecisive* model. For this problem, we present an approximation algorithm with factor $(1+O(\epsilon))$ and running time $O(2^{\frac{1}{\epsilon^d}}.\epsilon^{-2d}.n^3)$. We follow the idea introduced by Consuegra et al. in [3].

In section 3, we studied the *MinDiam* problem, where each imprecise point is represented with a convex regions. For this problem, we presented a polynomial time $\sqrt{d}$-approximation algorithm for constant $d$. next, we used this result to give a $(1+\epsilon)$-approximation for the case, where $d=2$ and the regions are $\alpha$-separate.

A future direction would be generalizing the $(1+\epsilon)$-approximation algorithm proposed for *MinDiam* to higher dimensions. In addition, one can think of removing the condition of $\alpha$-separability for this problem.